\documentclass[a4paper,aps,pra,twocolumn,showpacs,preprintnumbers,amsmath,amssymb]{revtex4}

\usepackage{bbm}
\usepackage{amsmath}
\usepackage{verbatim}
\usepackage{graphicx}
\usepackage{epsfig}

\def\ket#1{\left| #1\right>}
\def\bra#1{\left< #1\right|}

\def\be{\begin{equation}}
\def\ee{\end{equation}}
\def\bi{\begin{itemize}}
\def\ei{\end{itemize}}
\def\bea{\begin{eqnarray}}
\def\eea{\end{eqnarray}}
\def\bma{\begin{mathletters}}
\def\ema{\end{mathletters}}

\def\C{\hbox{$\mit I$\kern-.7em$\mit C$}}

\begin{document}
\title{On preparation of the W-states from atomic ensembles}

\author{V.N. Gorbachev\footnote{E-mail: vn@vg3025.spb.edu},
A.A. Rodichkina \footnote{E-mail: anastasia\_rodichkina@yahoo.de
}, A.I. Trubilko.}

\affiliation{Laboratory of Quantum Information and Computing,
University of AeroSpace Instrumentation, 67, Bolshaya Morskaya,
St.-Petersburg, 190000, Russia}

\begin{abstract}
A scheme, where three atomic ensembles can be prepared in the
states of the W-class via Raman type interaction of strong
classical field and a projection measurement involved three
single-photon detectors and two beamsplitters, are considered. The
obtained atomic entanglement consists of the Dicke or W-states of
each of the ensembles.
\end{abstract}

\pacs{03.67.-a}

\maketitle

\section{Introduction}
An atom of lambda-configuration seems to be attractive with
respect to state preparation, because one of its transitions
allows us to spy upon the atom behavior, say, by detecting of the
emitted photons. This is projection measurement, that is a way to
obtained a quantum system in some state, particulary entangled. In
fact, the entanglement generated through the Bell-state
measurement on two atomic samples has been demonstrated               
experimentally \cite{Pol}.\\ We focus on the states from the
W-class, introduced in ref. \cite{WCirac}, where some particular
members and their generalizations belong to family of the
symmetric Dicke states \cite{Dickest}. They are interesting
because of its robustness with respect to loss of a particle
\cite{Dicke}. In ref. \cite{PXGCG} a set of projection
measurements has been proposed to produce the W-state of atomic
ensembles, that consist of lambda-atoms. The W-state, generated in
this way, has been
considered for demonstrating Bell theorem \cite{Gou}. \\
In this paper we present a scheme for preparing the W- and
symmetric Dicke states. It contrast to scheme discussed in ref.
\cite{PXGCG}, it is more simple because of the introduced
projection measurement, that doesn't involve preparation of EPR
pair as an intermediate step, for example. Indeed, the obtained
W-entanglement of atomic ensembles is represented by the W-states
of atoms in each ensemble and can be particulary suitable as the
quantum channel for teleportation \cite{chanW}. Also during
evolution there are  some entangled states of light to be close to
ZSA (Zero Sum Amplitude) states introduced in ref. \cite{Pati}.

 Our paper is organized as follows.
First, we introduce hamiltonian with local field operators to
describe interaction of light with atomic ensembles spatially
separated. Next, we consider a family of W-states obtained while
evolution and projection measurement.

\section{Hamiltonian}
To describe interaction of light with atomic ensembles spatially
separated, we consider a hamiltonian, in which local field
operators are introduced. In the dipole approximation it has the
form \cite{Ham}
\begin{eqnarray}\label{10}
\nonumber
H=-i\hbar^{-1}\vartheta, &&\\
\nonumber \vartheta=\sum_{\mathbf{m},\mathbf{l}}
\sqrt{\frac{\hbar\omega_{m}}{2 \varepsilon_{0}a^{3}}}
A_{\mathbf{ml}}\exp(-i\omega_{m}t+i\mathbf{ml})d_{\bf{l}}&&\\
-h.c.&&
\end{eqnarray}
In (\ref{10}) local field  operators are represented by packets of
plain waves
\begin{equation} \label{11}
A_{\mathbf{ml}}=\frac{1}{\sqrt{M}}\sum_{k\sim
m}a_{k}\exp(-i(\omega_{k}-\omega_{m})t+i(\bf{k}-\bf{m})\bf{l}),
\end{equation}
which wave vectors $\mathbf{k}\sim \mathbf{m} $ lie in a band
$m_{s}-\pi/a\leq k_{s}<m_{s}+\pi/a $, $\Delta k_{s}=2\pi/L$,
$s=x,y,z$, where $L^{3}$ is a normalized volume, which modes are
described by operators $a_{\mathbf{k}}, a^{\dagger}_{\mathbf{k}}$,
$[a_{\mathbf{k}};a^{\dagger}_{\mathbf{k}'}]=\delta_{\mathbf{kk}'}$.
Packets are localized in a space sell $a^{3}=(L/M)^{3}$, which
position is given by vector $\mathbf{l}$. The local field
operators obey commutation relations $[A_{\mathbf{ml}},
A_{\mathbf{m'l'}}^{\dagger}]=\delta_{\mathbf{mm'}}\delta_{\mathbf{ll}'}$
and describe creation and annihilation of photon in a point
$\mathbf{l}$. In hamiltonian of interaction (\ref{10}) atoms are
presented by their dipole momentum operator $d_{\textbf{l}}$, that
is a total momentum of all atoms inside a sell. Assuming for
simplicity, that there is only one atom in any of the $M$ space
cells, one finds, that $d_{\textbf{l}}$ is the dipole momentum of
single atom, located in $\mathbf{l}$ and summing over $\textbf{l}$ is
summing over atoms.

Consider three atomic ensembles $A, B, C$ spatially separated and
consisted of the $\Lambda$-atoms with a ground  level $0$ and
exited levels $1$ and $2$ as it is shown in fig.1. Assuming a
Raman type interaction of strong classical field with atomic
transition $0-2$, that results in week waves scattered from $2-1$.
We shall describe light scattered using the introduced local field
operators. Then hamiltonian of interaction can be obtained from
(\ref{10}), where summation over $\textbf{l}$,  or atoms, is
rewritten as a sum over ensembles and a sum over atoms inside of
each ensemble. We shall neglect spatial behavior of all fields
inside any ensemble, assuming, that light interacts with such
family of atoms as a whole. Then hamiltonian takes the form
\begin{eqnarray}\label{13}
\nonumber
&&H=-i\hbar^{-1}\sum_{x=A,B,C}\vartheta_{x},\\
\nonumber
&&\vartheta_{x}=\Omega(S_{20}(x)\exp(-i(\omega-\omega_{20})t
+i\mathbf{kl}_{x})
-h.c.)\\
\nonumber &&+\sum_{\mathbf{m}}(\epsilon_{m}A_{\mathbf{m}x}\exp
(-i(\omega_{m}-\omega_{21})t+im\mathbf{l}_{x})S_{21}(x)\\
&&-h.c.),
\end{eqnarray}
where $\Omega$ is normalized amplitude of the classical field of
frequency $\omega$ and wave vector $\textbf{k}$,
$\epsilon_{m}=d\sqrt{\hbar\omega_{m}/2 \varepsilon_{0}a^{3}}$, $d$
is a dipole momentum, $S_{pq}(x)=\sum_{a\in x}|p\rangle_{a}\langle
q|$, $p,q=0,1,2$ is atomic operator of the ensemble $x$, which
position is $\textbf{l}_{x}$. In (\ref{13}), local operators
$A_{\mathbf{m}x}, A_{\mathbf{m}x}^{\dagger}$ describe creation and
annihilation of photon from the ensemble $x$ or in point $l_{x}$,
where $\bf{m}$ is the photon wave vector. The obtained Hamiltonian
allows to consider evolution of atoms and light scattered.

For simplicity we will take into account only single mode toward
resonance scattering. It results in one-dimension problem, where
$\omega\approx \omega_{20}$, $\omega_{m}\approx \omega_{12}$ and
$k\approx m$. In this approximation hamiltonian (\ref{13}) reads
\begin{eqnarray}\label{14}
\nonumber
\vartheta_{x}=\Omega(S_{20}(x)\exp(ikl_{x}) -h.c.)&&\\
+ \epsilon (A_{x}\exp(ikl_{x})S_{21}(x)- h.c.).&&
\end{eqnarray}
Let the initial total state be a vacuum in a sense that all atoms
are in its ground levels and all photons are in vacuum state
\begin{equation}\label{21}
\Psi_{0}=|000\rangle_{ABC}\otimes |0\rangle_{f},
\end{equation}
where $|000\rangle_{ABC}=|0\rangle_{A}\otimes|0\rangle_{B}
\otimes|0\rangle_{C}$,
$|0\rangle_{x}=|0\rangle\otimes\dots\otimes|0\rangle=
|0\rangle^{\otimes N_{x}}$, $N_{x}$ is total number of atoms of an
ensemble $x=A,B,C$, $|0\rangle_{f}$ is the field vacuum.

\section{W-states}

Collection of the light Fock states $\{\ket{n_{1}n_{2}n_{3}}\}$ is
a complete set and suitable to describe a measurement including
three single-photon detectors. For considering evolution of wave
function due from hamiltonian (\ref{14}), we will use this basis,
but we focus on three terms of the expansion, which are
represented by one and two-photon states. They are
$\ket{p_{x}}=\{\ket{p00}, \ket{0p0}, \ket{00p}\}$, $p=1,2$ and
$\ket{1_{x},1_{y}}= \{\ket{110}, \ket{101}, \ket{011}\}$, where
$x,y=A,B,C$. For example, if $x=A, y=B$ one finds single photon in
position of ensemble $A$: $\ket{1_{A}}=\ket{100}$, or two photons
in $A$ and $B$: $\ket{1_{A},1_{B}}=\ket{110}$ or two photons in
$A$: $\ket{200}$. Using the basis of Fock states we write
\begin{eqnarray}\label{210}
\Psi(t)=S(t)\Psi_{0}=\Psi_{\{1\}}+\Psi_{\{1,1\}}+\Psi_{\{2\}}+\dots,
&&
\end{eqnarray}
where $S(t)=\exp(-t\hbar^{-2}\sum_{x}\vartheta_{x})$ and  waves
functions $\Psi_{\{p\}}$, $p=1,2$,  $\Psi_{\{1,1\}}$, that
describe one photon and two photons emitted from atomic ensembles
read
\begin{eqnarray}\label{221}
\Psi_{\{p\}}=\sum_{x=A,B,C}\ket{p_{x}}\Psi(p_{x})&&
\end{eqnarray}
and
\begin{eqnarray}\label{222}
\Psi_{\{1,1\}}=\sum_{x,y=A,B,C}\ket{1_{x},1_{y}}\Psi(1_{x},1_{y}).&&
\end{eqnarray}

Using the theory of perturbation over interaction in the first
non-vanishing order coefficients in $\Psi_{\{1\}}$ take the form
\begin{eqnarray}\label{223}
\nonumber
 \Psi(1_{x})=\bra{1_{x}}S\ket{\Psi_{0}}&&\\=
-(1/2)(t\hbar^{-2})^{2}\Omega\epsilon
\bra{1_{x}}A_{x}^{\dagger}S_{10}(x)\ket{\Psi_{0}}.&&
\end{eqnarray}
Also we find
\begin{eqnarray}\label{224}
\nonumber
 \Psi(1_{x},1_{y})=\bra{1_{x},1_{y}}S\ket{\Psi_{0}}&&\\=
(3/4!)(t\hbar^{-2})^{4}(\Omega\epsilon)^{2}
A_{x}^{\dagger}A_{y}^{\dagger}S_{10}(x)S_{10}(y)\ket{\Psi_{0}}&&
\end{eqnarray}
and
\begin{eqnarray}\label{2241}
\nonumber
 \Psi(2_{x})=\bra{2_{x}}S\ket{\Psi_{0}}&&\\=
(3/4!)(t\hbar^{-2})^{4}(\Omega\epsilon)^{2} \bra{2_{x}}
A_{x}^{\dagger 2}S_{10}(x)S_{10}(x)\ket{\Psi_{0}}.&&
\end{eqnarray}

Two points maybe made about it. First, any atomic states
$S_{e0}^{m}(x)|0\rangle_{x}$,  $e=1,2$, $m=1,\dots, N_{x}$ are
known as Dicke states, described $m$ excited atoms from $N_{x}$.
They read
\begin{equation}\label{24D}
S_{10}^{m}\ket{0}=m!\sqrt{C_{N,m}}\ket{m,N},
\end{equation}
where we omit the ensemble index $x$, and the normalized ket is
introduce
\begin{equation}\label{24DN}
\ket{m,N}=1/\sqrt{C_{N,m}}\sum_{z}P_{z}(\ket{1}^{\otimes
m}\otimes\ket{0}^{\otimes (N-m)}),
\end{equation}
where $P_{z}$ is set of $C_{N,m}=N!/m!(N-m)!$ distinguished
permutations of particles. For particular case $m=1,2$ one finds
the states of the W-class
\begin{eqnarray}\label{24W1}
\nonumber
 W_{1}\equiv\ket{1,N}&&\\
=\frac{1}{\sqrt{N}} (|10\dots 0\rangle+|01\dots 0\rangle+
\dots+|00\dots 1\rangle),&&
\\
\label{24W2} \nonumber
 W_{2}
\equiv\ket{2,N}&&\\
\nonumber =\sqrt{\frac{2}{N(N-1)}} (|110\dots
0\rangle+ 
+|101\dots 0\rangle+ \dots &&\\
+ |0\dots 011\rangle).&&
\end{eqnarray}

Second, in Eq. (\ref{223}), (\ref{224}) and (\ref{2241}) the Fock
states of light $A_{x}^{\dagger}|0\rangle_{f}$,
$A_{x}^{\dagger}A_{y}^{\dagger}\ket{0}_{f}$
 and $A_{x}^{\dagger 2}\ket{0}_{f}$ are localized states field, that
 describe one photon, emitted from atomic ensemble $x$,
 two photons each of which from $x$ and $y$, and a pair of quanta
 from $x$. Indeed, they look as W-states. The reason is that in
accordance with (\ref{11}) the local operator $A_{x}$ is a
superposition of $M$ operators $a_{k}$
\begin{eqnarray}\label{26}
&& A_{x}^{\dagger}\ket{0}=\exp(iml_{x})\ket{\tilde{W}},
\end{eqnarray}
where
\begin{eqnarray}\label{27}
\nonumber &&\ket{\tilde{W}}=
 \frac{1}{\sqrt{M}}\sum_{k\sim
m}\exp(-ikl_{x})a_{k}^{\dagger}\ket{0}\\
&&=e^{i\zeta}\sum_{q=0}^{M-1} C_{q}\ket{0}^{\otimes
q}\ket{1}\ket{0}^{\otimes (M-q-1)}.
\end{eqnarray}
In Eq (\ref{27}) the phase factor and coefficients read
$\zeta=(\pi/a-m)l_{x}$, and $C_{q}=\exp(-i2\pi
ql_{x}/L)/\sqrt{M}$. Does the state $\ket{\tilde{W}}$ belong to
the W-class? Note, if $M\gg 1$, then in (\ref{27}) sum of
coefficients $C_{q}$ is
\begin{eqnarray}\label{28}
\nonumber && \sum_{q}C_{q} \propto \exp(iy)\sin(y)/y, \quad y=\pi
l_{x}/a.
\end{eqnarray}
This sum is equal to zero, if $y\neq 0$, then maybe it is a
reason, that the considered state is close to the ZSA-states,
introduced in ref. \cite{Pati}.

As result the $\Psi_{\{1\}}$ takes the form
\begin{eqnarray}\label{225}
\nonumber
\Psi_{\{1\}}=-(t^{2}\hbar^{-4}\Omega\epsilon/2)(\sqrt{N_{A}}\ket{100}_{f}
\otimes\ket{W_{1}00}&&\\
+\sqrt{N_{B}}\ket{010}_{f} \otimes\ket{0W_{1}0}+
\sqrt{N_{C}}\ket{001}_{f} \otimes\ket{00W_{1}}),&&
\end{eqnarray}
where $\ket{W_{1}}$ is given by (\ref{24W1}). This is a
superposition state that is obtained after a photon is emitted
from one of the three ensembles. Equation (\ref{225}) tells, if an
ensemble emits one photon, then their atoms are prepared in the
W-state. Also, when two photons arise from two ensembles, one
concludes, that both ensembles are in the W-state.  It  is
described by the wave function $\Psi_{\{1,1\}}$, that has the form
\begin{eqnarray}\label{226}
\nonumber \nonumber
 \Psi_{\{1,1\}}=&&\\
\nonumber
 =(t^{2}\hbar^{-4}\Omega\epsilon)^{2}6/4!)
(\sqrt{N_{A}N_{B}}\ket{110}_{f}
\otimes\ket{W_{1}W_{1}0}&&\\
\nonumber
 +\sqrt{N_{A}N_{C}}\ket{101}_{f}
\otimes\ket{W_{1}0W_{1}}&&\\+ \sqrt{N_{B}N_{C}}\ket{011}_{f}
\otimes\ket{0W_{1}W_{1}}).&&
\end{eqnarray}
Atomic states of the $W_{2}$ type are presented in
$\Psi_{\{2\}}$, that reads
\begin{eqnarray}\label{2261}
\nonumber \nonumber
 \Psi_{\{2\}}=&&\\
\nonumber
 =(t^{2}\hbar^{-4}\Omega\epsilon)^{2}3/4!)
(\sqrt{N_{A}(N_{A}-1)}\ket{200}_{f}
\otimes\ket{W_{2}00}&&\\
\nonumber
 +\sqrt{N_{B}(N_{B}-1)}\ket{020}_{f}
\otimes\ket{0W_{2}0}&&\\+ \sqrt{N_{C}(N_{C}-1)}\ket{002}_{f}
\otimes\ket{00W_{2}}),&&
\end{eqnarray}
where $\ket{W_{2}}$ is given by (\ref{24W2}).

\section{Projection measurement}

Assuming a measurement performing on light scattered by atoms,
where three detectors $D_{x}$, $x=A,B,C$ collect photons, that
come from ensembles $x$, so that detector $D_{A}$ can't see any
photons from $B$ and $C$ etc. It follows from (\ref{225}), that
for example, the state $\ket{W_{1}00}$ can be prepared, when
outcome of the measurement is $\ket{100}$, or there is a click of
the detector $D_{A}$.  This state, achieved with probability
$Prob(100)=(t^{2}\hbar^{-4}\Omega\epsilon/2)^{2}N_{A}$, is
separable with respect to atomic ensembles. However a
three-ensemble entanglement can be also prepared.

Consider a detection scheme, shown in fig.1. It includes two
beamsplitters $BS_{1}$ and $BS_{2}$ with transmittances and
reflectances $c_{k}, s_{k}$, $c_{k}^{2}+s^{2}_{k}=1$, $k=1,2$.
\begin{figure}[ht]
  \centering
\epsfxsize=5cm \epsfbox{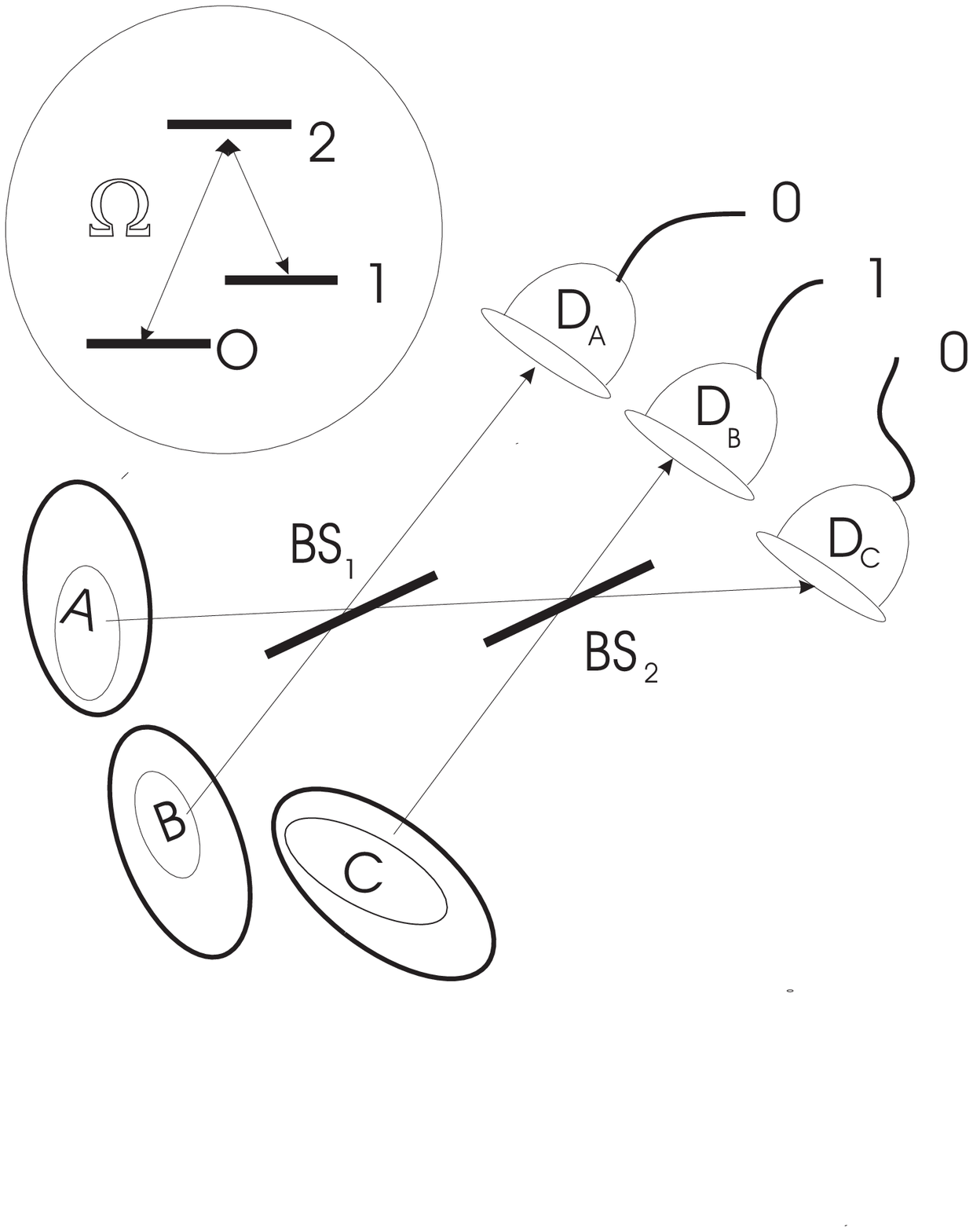}
  \caption{Three ensembles of the $\Lambda$- atoms are prepared in a state of W-class by
projection measurement. Atoms are excited by a strong light at
frequency $\Omega$, then the week light scattered from transition
2$\to$1 is mixed by beamsplitters $BS_{1}$ and $BS_{2}$ and comes
to detectors $D_{A}, D_{B}, D_{C}$.}\label{1}
\end{figure}
Let photons coming from the $A, B$ and $C$ ensembles are mixed by
these beamsplitters  before  measuring, when photon from $A$ and
$B$ are inputs of the first beamsplitter and the photon from $C$
is mixed with one of the outputs of the first beamsplitter.
Assuming, that  beamsplitter can be described by Hamiltonian
$H=i\hbar k(a^{\dagger}b-ab^{\dagger})$, where $a$ and $b$ are
photonic operators, $k$ is a coupling constant, we find, that
$BS_{1}$ and $BS_{2}$ transform the single-photon Fock-state as
\begin{eqnarray}\label{31}
\nonumber
  a\ket{100}+b\ket{010}+c\ket{001}\to&&\\
\nonumber
  \phi=a[c_{1}\ket{100}-s_{1}(c_{2}\ket{010}+s_{2}\ket{001})]&&\\
  \nonumber
  +b[s_{1}\ket{100}+c_{1}(c_{2}\ket{010}-s_{2}\ket{001})]&&\\
  +c[s_{2}\ket{010}+c_{2}\ket{001})].&&
\end{eqnarray}
Indeed, if $b=c=0$ outgoing is a state of the W-class. When
$c_{1}=-s_{1}=1/\sqrt{2}$ and $c_{2}=\sqrt{2/3}$ output state of
light to be measured by detectors  takes the form
\begin{eqnarray}\label{32}
\nonumber \phi=1/\sqrt{2}(a-b)\ket{100}
  &&\\
  \nonumber
+1/\sqrt{3}(a+b+c)\ket{010}
  &&\\
+(-a/\sqrt{6}-b/\sqrt{6}+c\sqrt{2/3})\ket{001}.&&
\end{eqnarray}
Using this transformation, one finds
\begin{eqnarray}\label{227}
\nonumber \Psi_{\{1\}}\to
\Psi_{\{1\}}'=-(t^{2}\hbar^{-4}\Omega\epsilon/2)&&\\
\nonumber (\ket{100}_{f}
\otimes(\sqrt{N_{A}}\ket{W_{1}00}-\sqrt{N_{B}}\ket{0W_{1}0})/\sqrt{2}
&&\\
\nonumber +\ket{010}_{f}\otimes
(\sqrt{N_{A}}\ket{W_{1}00}+\sqrt{N_{B}}\ket{0W_{1}0}&&\\
\nonumber
 + \sqrt{N_{C}}\ket{00W_{1}})/\sqrt{3}
&&\\
\nonumber
 +\ket{001}_{f}\otimes
(-\sqrt{N_{A}}\ket{W_{1}00}-\sqrt{N_{B}}\ket{0W_{1}0}&&\\
+ 2\sqrt{N_{C}}\ket{00W_{1}})/\sqrt{6}).&&
\end{eqnarray}
It follows from (\ref{227}), that the states of EPR and W-class
can be prepared. Suppose, any of ensembles has the same number of
atoms $N$. When outcome $\ket{100}$ or $\ket{010}$ is obtained,
atomic ensembles are prepared in the states
$(\ket{W_{1}00}-\ket{0W_{1}0})/\sqrt{2}$ and
$(\ket{W_{1}00}+\ket{0W_{1}0}+\ket{00W_{1}})/\sqrt{3}$. It can be
done with probability
$Prob(100)=Prob(010)=(t^{2}\hbar^{-4}\Omega\epsilon/2)^{2}N$. The
similar way results in the W-state of the form
$(\ket{W_{1}W_{1}0}+\ket{W_{1}0W_{1}}+\ket{0W_{1}W_{1}})/\sqrt{3}$.
It needs $c_{1}=s_{1}=c_{2}=s_{2}$ and outcome to be $\ket{101}$.

Manipulating number of atoms or another parameter a state of the
form $ 1/\sqrt{2}(\ket{W_{1}00}+\sqrt{2}\ket{0}\Psi^{+})$, where
$\Psi^{+}=(\ket{0W_{1}}+\ket{W_{1}0})/\sqrt{2}$, can be achieved.
This three-partite entanglement of the W-class is unitary
equivalent to the state of the GHZ-class and hence it is
sufficient as a quantum channel for perfect teleportation
\cite{chanW}.

\begin{acknowledgments}
We acknowledge stimulating discussions with A. Basharov. This work
was supported in part by the Delzell Foundation Inc. and INTAS
grant no. 00-479.
\end{acknowledgments}


\end{document}